\documentclass[aps,showpacs,nofootinbib]{revtex4}
\usepackage{amsmath}
\usepackage{amsfonts}
\usepackage{amssymb,amscd,amsbsy}
\usepackage{epsfig}
\usepackage{dsfont}
\usepackage{color}

\newcommand{\bea}{\begin{eqnarray}}
\newcommand{\eea}{\end{eqnarray}}

\newcommand{\vect}[1]{\mathbf{#1}}

\newcommand{\di}{\displaystyle}
\newcommand{\req}{\rho_{\rm eq}}

\newcommand{\kt}{k_{\rm B}T}
\newcommand{\rref}{\rho_{\rm ref}}
\newcommand{\cref}{c^{(2)}_{\rm ref}}
\newcommand{\tcref}{\tilde c^{(2)}_{\rm ref}}

\begin{document}

\title{Free energies, vacancy concentrations and density distribution anisotropies in 
 hard--sphere crystals: A combined density functional and simulation study} 

\author{M. Oettel$^{1,2}$, S. G\"orig$^1$, A. H\"artel$^3$, {H. L\"owen$^3$, }M. Radu$^{1,4}$, and
  T. Schilling$^4$ }
\affiliation{ $^1$ Johannes Gutenberg--Universit\"at Mainz, Institut f{\"ur} Physik,
  WA 331, D--55099 Mainz, Germany \\
              $^2$ \textit{Material- und Prozesssimulation,
Universit{\"a}t Bayreuth, N{\"u}rnberger Stra{\ss}e 38, D--95448 Bayreuth, Germany} \\
              $^3$ Institut f\"ur Theoretische Physik II: Weiche Materie,
  Heinrich--Heine--Universit\"at D\"usseldorf,
  Universit\"atsstra{\ss}e 1, D--40225 D\"usseldorf, Germany \\
              $^4$ Universit\'e du Luxembourg, Theory of Soft Condensed Matter, L-1511 Luxembourg, Luxembourg  
}


\begin{abstract}
We perform a comparative study of the free energies and the 
density distributions in hard sphere crystals using Monte Carlo simulations and 
density functional theory (employing Fundamental Measure functionals). 
Using a recently introduced technique (Schilling and Schmid, J.~Chem.~Phys {\bf 131}, 231102 (2009))
 we obtain crystal free energies
to a high precision.
The free energies from Fundamental Measure theory are in good agreement with 
the simulation results and demonstrate the applicability of these functionals 
to the treatment of other problems involving crystallization. The agreement between
FMT and simulations on the level of the free energies is also reflected in the 
density distributions around single lattice sites.
Overall, the peak widths and anisotropy signs for different
lattice directions agree, however, it is found that  Fundamental Measure theory 
 gives slightly narrower peaks with more
anisotropy than seen in the simulations.   
Among the three types of Fundamental Measure functionals studied, only the White Bear II
functional (Hansen--Goos and Roth, J.~Phys.: Condens.~Matter {\bf 18}, 8413 (2006))
exhibits sensible results for the equilibrium vacancy concentration and a physical behavior
of the chemical potential in crystals constrained by a fixed vacancy concentration.
\end{abstract}

\pacs{82.70Dd,61.50Ah,71.15Mb}

\maketitle

\section{Introduction}

The phase behavior of hard spheres {is} one of the most intensely studied
subjects within the realm of classical statistical mechanics. The existence
of a fluid--solid transition has been predicted already more than fifty years ago
by early computer simulation methods \cite{Woo57,Ald57}. {A}dvances in colloidal engineering
have led to the experimental realization of almost hard sphere--like systems,
confirming the occurence of crystallization in such systems in the 1980's \cite{Pus86}. The variety
of hard sphere--like colloidal systems include polymeric spheres 
(index-matched solvent, sterically stabilized) \cite{Bry02} and also thermotropic colloids \cite{Sie09}
using particles with diameters of the order of a few hundred
nm. This allows the use of scattering techniques with visible light and/or the use {of} 
real--space microscopy to resolve single--particle positions. 
Using these systems and techniques, numerous features of the statics and dynamics of the 
crystallization process 
and the competing glass transition have been studied in detail (see e.g. Refs.~\cite{Gas01,Wee02, Scho06, Zac09, Sch10}).

The progress in real--space imaging opens the perspective that the static
density distribution in crystals and the dynamics of the nucleation process can 
be studied with unprecedented resolution. The primary information obtained
in these experiments, the trajectories of single particles, is very much the same
as the information obtained in a computer simulation. Thus
the further analysis of this primary information brings together these two fields.
Currently, e.g. the processes of homogeneous and heterogeneous nucleation
in colloidal hard sphere systems are under scrutiny {\cite{Emm09,Kah09}}, and the unambiguous 
resolution of the underlying mechanisms of these processes appears to be possible
using simulation/real--space experiment on the one side and the established
reciprocal--space (scattering) experiments on the other. 

From the theory side, classical density functional theory (DFT) is a good candidate to study 
crystallization phenomena on a microscopic level. The concepts of equilibrium DFT     
have been developed over the past forty years (for an early review see Ref.~\cite{Eva79}). 
In this context, hard spheres appear
to be {one of the few} classical fluid{s} for which quantitatively predictive 
density functionals can be constructed {thanks} to powerful geometric arguments, leading
to the so-called Fundamental Measure Theory (for recent reviews see Refs.~\cite{Tar08,Rot10}). 
In contrast to the maturity of
equilibrium DFT, dynamic DFT is a still developing field which has
 has been started only about ten years ago \cite{Mar99,Arc04,Arc09,Esp09,Rex08}.  
Centerpiece of dynamic DFT is the time evolution of the inhomogeneous one--particle
density. The most
intensely studied variant of the theory is actually an approximation to Brownian
dynamics, thus it appears to be well--suited for the study of colloidal systems. 
However, due to the complexity of the FMT functionals, any dynamic DFT studies 
of a hard sphere system with inhomogeneities in two or three dimensions have not been
undertaken. In fact, there are only a few equilibrium studies of inhomogeneous 
problems in two and three dimensions \cite{Fri00,Gou01,Oet09,Bot09}, unrelated to the crystallization problem.
The study of hard sphere crystals within FMT has been restricted so far to 
sensible parametrizations of the density distribution in a crystal,
{nevertheless this approach has elucidated}  the key features of a reliable DFT for the 
crystallization
transition \cite{Tar00}. (A more detailed review of the problem of crystal phases within
density functional theory is given below.)  
  
In order to make progress in the direction of applying dynamic DFT (with the 
FMT functionals {that work very well} for hard spheres) to the currently studied nucleation 
problems {\cite{Emm09,Kah09}},
we will study first the more modest problem of {the} static density distribution in hard
sphere crystals in this paper. This will be done by a full, three--dimensional minimization  
of the FMT functionals and contrasted to the results of our Monte--Carlo simulations.
(Surprisingly, the density distribution in hard sphere crystals has been likewise
studied very little using simulations.) Such a study is an absolute 
prerequisite for the more difficult dynamic problems involving crystal--fluid
interfaces to be tackled in the future. 
We will show that the full minimization discriminates 
between different FMT functionals which are very similar in the description of
the fluid phase. We will shed some {new} light on the problem of an equilibrium 
vacancy concentration within DFT. We will demonstrate that the FMT results for the
free energy per particle for the crystal phase are in very good agreement with
the corresponding simulation result {which has been produced by a recently introduced method.}   

The paper is structured as follows. In Sec.~\ref{sec:dft} we briefly review the
density functional approach to crystallization in the hard sphere system.
Sec.~\ref{sec:fmt} discusses a few points relevant for the FMT crystal 
description {in} more depth. These address the constrained minimization 
in the unit cell (with particle 
number fixed), the relation of
the respective constrained chemical potential to Widom's trick in a system with
fixed vacancy concentration, and the numerical 
procedure of the FMT functional minimization. 
In Sec.~\ref{sec:mc} we briefly describe our Monte Carlo {method} to obtain free energies
and density distributions. Sec.~\ref{sec:results} {compiles} our results on free energies,
equilibrium vacancy concentrations and density distributions and  in Sec.~\ref{sec:conclusions}
we present our conclusions.

\section{Hard sphere crystals in density functional theory}

\label{sec:dft}

In density functional theory, the crystal is viewed as a self--sustained inhomogeneous
fluid, i.e. an inhomogeneous density profile $\rho_{\rm cr}(\vect r)$ minimizes the
grand potential functional
\bea
 \Omega [\rho(\vect r)] &=& {\cal F}[\rho(\vect r)] - \int d^3r \rho(\vect r)
  \left( \mu - V^{\rm ext}(\vect r)\right) \;,
\eea
with the external potential $V^{\rm ext}$ being zero. Here, $\mu$ is
the chemical potential and  ${\cal F}[\rho]$ is the
free energy functional which is conventionally split into an ideal and an excess
part:
\bea
  {\cal F}[\rho] &=& {\cal F}^{\rm id} [\rho] +  {\cal F}^{\rm ex}[\rho]
\eea
with the exact form of the ideal part given by
\bea
  {\cal F}^{\rm id}[\rho] &=& \int d^3r\,f^{\rm id}(\vect r) = \int d^3r\, \rho(\vect r )\left( 
 \ln[ \rho(\vect r) \Lambda^3] -1 \right)\;.
\eea
Here, $\Lambda$ is the de--Broglie wavelength.
It was realized very early (in 1979) that a simple Taylor--expanded version of the
excess free energy
\bea
 \label{eq:fhnc}
   {\cal F}^{\rm ex}[\rho] &= & \int d^3r\,f^{\rm ex}(\vect r) \approx  
   - \frac{1}{2} \int d^3r \int d^3r' \cref(\vect r - \vect r'; \rref)
   \Delta\rho(\vect r) \Delta\rho(\vect r') 
\eea
allows for {minimizing} solutions $\rho_{\rm cr}(\vect r)$ \cite{Ram79}. Here,
$\rref$ is a reference density around the liquid coexistence density and
$\cref$ is the direct correlation function in the bulk liquid at this reference
density (which is related to the bulk structure factor by
$S(k) = 1/\rref - 1/\tcref(k)$). 
In this early work, the minimization to obtain $\rho_{\rm cr}$ was a constrained one:
expanding the density as
\bea
 \label{eq:rhoexpansion}
  \rho_{\rm cr}(\vect r) = \rho_0 + \sum_j \rho_j \exp( {\rm i} \vect K_j \cdot \vect r)
\eea 
($\vect K_j$ is the set of the reciprocal lattice vectors), the minimization
was only performed with respect to the moments $\rho_j$ which belong to the first
or to the first and fourth shell of reciprocal lattice vectors.\footnote{Reciprocal lattice 
vectors belong to the same shell if they transform into
each other under the point group transformations from the considered crystal symmetry.
The first shell contains all reciprocal lattice vectors with the lowest magnitude,
etc. As an example, for fcc, the reciprocal lattice is bcc and the first shell contains 8 reciprocal lattice vectors
$2\pi/a (\pm 1, \pm 1, \pm 1)$ where $a$ is the side length of the cubic unit cell.
The fourth shell (used in Ref.~\cite{Ram79}) contains 24 reciprocal lattice vectors, given by
$2\pi/a (\pm 3, \pm 1, \pm 1)$ plus two cyclic permutations of the Cartesian components.
} 
 In this approximation, one sees
that the crystal free energy in Eq.~(\ref{eq:fhnc}) ``probes'' the Fourier
transform $\tcref(k)$ only at one or two values of $k$. 
For fcc these
values are $k_1 \approx 10.9/a$ and $k_4 \approx 20.8/a$  where $a$ is the
side length of the cubic unit cell, very near the first two maxima of $\tcref(k)$ resp.
$S(k)$. At first sight, it may appear surprising that an expansion of the free energy
like Eq.~(\ref{eq:fhnc}), valid at {\em small} density variations is sufficient
to sustain the rapidly varying density profile in a crystal. 
However, the isotropic correlations between two particles in Fourier space 
are described by the structure factor (and hence by $\cref$).  Since the 
shells of reciprocal lattice vectors for an fcc lattice are also distributed 
fairly isotropically, the possible description of a solid with a density near the reference density
appears to be less unexpected. Subsequent work has revealed that the expansion
in reciprocal space (\ref{eq:rhoexpansion}) is converging slowly. Furthermore, 
there are serious quantitative problems in this
approach if it comes to the description of the lattice density peak width (much too narrow), 
crystals at higher density (unstable) and
the vacancy density (around 10 percent at coexistence which is a factor of about 100 too large) \cite{Hay85,Jon85}. 

A more general approach to inhomogeneous hard sphere fluids in general and to 
the description of
crystals in particular consists in the {\em ansatz}
\bea
 \label{eq:fwda}
 {\cal F}^{\rm ex}[\rho] = \int d^3 r\, \rho(r) \Psi( \bar \rho (\vect r))\;.
\eea 
Here, $\Psi$ is a suitable function of a weighted density 
\bea
  \bar \rho(\vect r) = \int d^3r' \rho(\vect r') w(\vect r- \vect r';\bar \rho) =
     \rho * w\,(\vect r)
\eea
which employs a weight function which in turn may depend on the weighted density itself.
(For the Taylor expanded functional (\ref{eq:fhnc}), $\bar \rho = {\rm const.} +
 \cref * \rho$.) The functions $\Psi$ and $w$ can be determined through the equation
of state and the bulk direct correlation functions which are assumed to be known.
Here, the self--consistent solution for $w$ may be rather involved in particular
realizations. 
Examples for this class of functionals include the Tarazona functionals 
Mark I \cite{Tar84} and Mark II \cite{Tar85}, the weighted--density approximation (WDA)
\cite{Cur85} and the modified WDA \cite{Den95}. Crystal structures in these approaches have been
usually obtained by minimizing the {\em ansatz}
\bea
  \label{eq:Gauss_ansatz}
  \rho_{\rm cr} &= & \sum_{{\rm lattice\;sites}\;i} N\,\left(\frac{\alpha}{\pi}\right) ^{\frac{3}{2}}
  \,\exp\left( -\alpha (\vect r - \vect r_i)^2
   \right)   
\eea    
with respect to the Gaussian peak width $\alpha$ and the normalization $N$. If $n_{\rm vac}$ denotes the
relative concentration of vacancies then $N=1-n_{\rm vac}$. With such a Gaussian {\em ansatz},
the reciprocal lattice modes of the density (see Eq.~(\ref{eq:rhoexpansion})) are given by     
$\rho_j = N\,\exp( - \vect K_j^2/(4\alpha))$. Using the most sophisticated versions of these 
weighted--density approaches, one can achieve a rather good agreement with simulations
for the liquid--solid coexisting densities and a physically sensible behavior also for denser crystals.
This is understandable since in comparison with the simple Taylor expanded functional
(\ref{eq:fhnc}) the weighted--density form (\ref{eq:fwda}) includes contributions from
higher--order direct correlation functions and through the self--consistent determination
of $w$ it is guaranteed that at higher densities the changed isotropic correlations in a (possibly
metastable) reference liquid are taken into account. Still, the lattice density peaks come out
too narrow compared to simulations, the crystal free energy per particle is too small
by about 5\% and a quasifree minimization in modified WDA (with the
restriction $n_{\rm vac}=0$) revealed qualitatively wrong peak asymmetries in the different
lattice directions as well as an unphysically large interstitial density \cite{Ohn93}.   
A sensible, small vacancy concentration $n_{\rm vac,0}$ which minimizes the free energy
can only be obtained by incorporating an appropriate additional constraint term into the
free energy functional \cite{Ohn91}.

\begin{figure}
  \begin{center}
    \epsfig{file=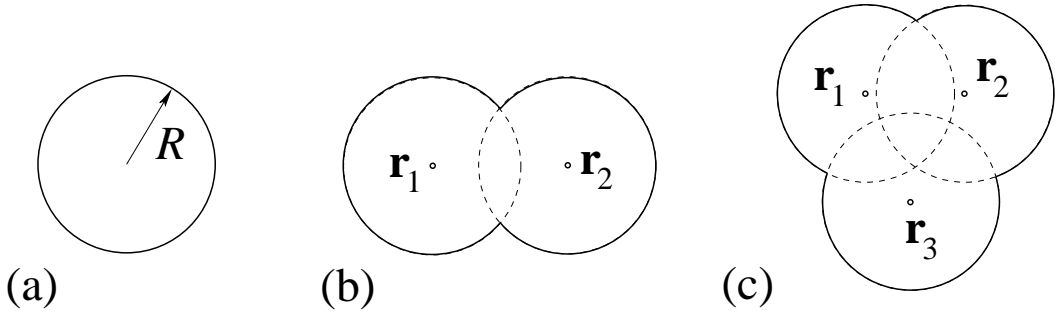, width=14cm}
    \caption{Three types of cavities which can hold only one particle.}
    \label{fig:cav}
  \end{center}
\end{figure}

However, the WDA approach which is built on the isotropic fluid correlations 
cannot be expected to treat coordination effects in crystals correctly on a fundamental level. 
These include 
the description of the metastable hard sphere bcc crystal \cite{Run87} and the crystal--fluid 
interface \cite{Ohn94}. Here, the development of fundamental measure theory (FMT) marks
an important breakthrough \cite{Ros89}. FMT postulates an excess free energy with a  local free energy 
density in a set of weighted densities $n_\alpha$:
\bea
  {\cal F}^{\rm ex}[\rho] &=& \int d^3r \Phi(n_\alpha(\vect r)) \;.
\eea   
The weighted densities are again constructed as convolutions of the density with weight functions,
$n_\alpha(\vect r) = \rho * w^\alpha(\vect r)$. In contrast to the WDA, 
the weight functions reflect
the geometric properties of the {\em individual} hard spheres (and not the properties of an
interacting pair). For one species, the weight functions include four scalar functions
$w^0 \dots w^3$, two vector functions $\vect w^0, \vect w^1$ and a tensor function
$w^t$ defined as  
\bea
 w^3 = \theta(R-|\vect r|)\;, \qquad
 w^2 = \delta(R-|\vect r|)\;, \qquad w^1 = \frac{w^2} {4\pi R}\;,
  \qquad w^0 = \frac{w^2}{4\pi R^2}\;,
  \nonumber\\
 \vect w^2 =\frac{\vect r}{|\vect r|}\delta(R-|\vect r|)\;,
 \qquad \vect w^1 = \frac{\vect w^2}{4\pi R} \;, \nonumber \\
 w^t_{ij} = \frac{r_i r_j}{\vect r^2} \delta(R-|\vect r|) \;.
\eea
Here, $R$ is the hard sphere radius. Using these weight functions,
corresponding scalar weighted densities $n_0\dots n_3$, vector weighted densities
$\vect n_1, \vect n_2$ and one tensor weighted density $n_t$ are defined. 
In constructing the free energy density $\Phi$, arguments
concerning the correlations in the bulk fluid and arguments for strongly inhomogeneous
systems are used. For the bulk, $\Phi$ is required to reproduce exactly the second and third
virial coefficent of the direct correlation function. Furthermore,
imposition of the Carnahan--Starling equation of state \cite{Rot02,Yu02} and/or
consistency with a scaled particle argument \cite{Ros89,Han06} leads
to a closed form for $\Phi$. The arguments using strongly inhomogeneous systems are known
in the literature under the label ``dimensional crossover'' \cite{Ros97}: Through a suitable external
potential, the hard sphere fluid can be confined to lower dimensions and the density functionals
for these lower--dimensional systems should emerge from the correct density functional in 3d.    
Of particular relevance are the crossover to 1d where the exact density functional is known 
\cite{Per76} and to 0d where a hard sphere is confined to a point and the free energy is a simple
function of the mean occupation number at this point \cite{Ros97}. The 0d confinement can be realized
through differently shaped cavities (overlapping spheres of radius $R$) which can hold only 
one particle (see Fig.~\ref{fig:cav}). Respecting the 0d limit for different cavities is of 
particular relevance for the crystal description since this means that the mutual exclusion of
hard spheres in various coordinations is correctly described. In Refs.~\cite{Tar97,Tar00} a solution
is given which respects the 0d limit for cavities (a) and (b) of Fig.~\ref{fig:cav} and approximates
the 0d limit for cavity (c). 

The arguments presented in the above paragraph lead to the following form of the
excess free energy density:
\bea
 \label{eq:phi_hs}
   \Phi( \{\vect n[\rho (\vect r)]\} ) &=&   -n_0\,\ln(1-n_3) +
      \varphi_1(n_3)\;\frac{n_1 n_2-\vect n_1 \cdot \vect n_2}{1-n_3} +
   \nonumber   \\
      & & \varphi_2(n_3)\; \frac{3 \left( -n_2\, \vect n_2\cdot \vect n_2 + n_{2,i} n_{t,ij} n_{2,j}
        + n_2\,  n_{t,ij}  n_{t,ji} - n_{t,ij}  n_{t,jk} n_{t,ki}
   \right) }{16\pi(1-n_3)^2}\;.
\eea
Here, $\varphi_1(n_3)$ and $\varphi_2(n_3)$ are functions of the local packing density $n_3(\vect r)$. 
With the choice
\bea
 \label{eq:frf}
 \varphi_1 = 1\;, \qquad \varphi_2 = 1
\eea
we obtain the Tarazona tensor functional \cite{Tar00} which is built upon the original
Rosenfeld functional \cite{Ros89}. The latter gives the fluid equation of state and
pair structure of the Percus--Yevick approximation. Upon setting
\bea
 \label{eq:fwb}
 \varphi_1 &= & 1 \\
 \varphi_2 & = & 1 - \frac{-2n_3+3n_3^2 - 2(1-n_3 )^2 \ln(1-n_3 ) }
                          {3 n_3^2}  \nonumber
\eea
we obtain the tensor version of the White Bear functional \cite{Rot02}, consistent with the quasi--exact
Carnahan--Starling equation of state. Finally, with
\bea
 \label{eq:fwbII}
 \varphi_1 & = & 1 + \frac{2n_3-n_3^2 + 2(1-n_3) \ln(1-n_3 )}{3n_3} \\
 \varphi_2 & = & 1 - \frac{2n_3-3n_3^2 + 2n_3^3 + 2(1-n_3 )^2 \ln(1-n_3 ) }
                          {3 n_3^2}  \nonumber
\eea
the tensor version of the recently introduced White Bear II functional \cite{Han06} is recovered.
This functional is most consistent with restrictions imposed by morphological thermodynamics
\cite{Koe04}.

\section{Hard sphere crystals in FMT}

\label{sec:fmt}

\subsection{Minimization and $\mu$ consistency}

\label{subsec:muconsistency}

The minimization of the grand potential functional
\bea
  \Omega[\rho] &=& {\cal F}^{\rm id}[\rho] +  {\cal F}^{\rm ex}[\rho] - \int
   d^3 r(\mu - V^{\rm ext}(\vect r)) \rho(\vect r) \;,
\eea
leads to
\bea
 \label{eq:rhoeq}
  \beta^{-1}\ln (\req(\vect r)\Lambda^3) =-\mu^{\rm ex} [\req(\vect r)] + \mu - V^{\rm ext}(\vect r) \;.
\eea
The functional $\mu [\rho(\vect r)]$ is given by
\bea
 \mu^{\rm ex} [\rho(\vect r)] &=&  \frac{\delta {\cal F}^{\rm ex}[\rho]}
      {\delta \rho(\vect r)} \\
\label{eq:mufunc}
    &=& \beta^{-1} \sum_\alpha \int d\vect r'\, \frac{\partial \Phi}{\partial n_\alpha(\vect r')}
         \,w^\alpha(\vect r'-\vect r) \;.
\eea
In principle, for a force--free system ($V^{\rm ext}=0$), the specification of a suitable
chemical potential $\mu$ should lead upon minimization to a periodic crystal profile
$\req \equiv \rho_{\rm cr}(\vect r)$ with the bulk density $\rho_0(\mu)$. The side length $a$ of the
cubic unit cell and consequently the vacancy concentration $n_{\rm vac}$ should adjust itself 
to comply with Eq.~(\ref{eq:rhoeq}). Here, $n_{\rm vac}$ is connected to the occupation of the
unit cell of the fcc lattice by
\bea
 \label{eq:nvacdef}
   \int_{\rm cell} d^3r\, \rho(\vect r) = 4 (1 - n_{\rm vac})\;.
\eea  
In practice, such a procedure is not feasible. Rather, for a given bulk density 
$\rho_0${,} also $n_{\rm vac}$ (and thus $a$) is prescribed and  
a constrained free energy functional for the
unit cell 
\bea
 \label{eq:fconstr}
 \left.\Omega'\right|_{\rm cell} = \int_{\rm cell} d^3r\, {f}^{\rm id}[\rho] +  
\int_{\rm cell}d^3r\, {f}^{\rm ex}[\rho] - \mu' \int_{\rm cell}d^3r\,  (\rho(\vect r)- \rho_0)
\eea
is minimized where $\mu'=\mu'(\rho_0,n_{\rm vac})$ plays the role of a Lagrange multiplier to ensure (\ref{eq:nvacdef}). 
In the work reviewed previously $\mu'$ was not determined explicitly, and only a few studies
bothered to vary also $n_{\rm vac}$ (which should be close to zero) 
such that the free energy per particle is indeed
minimized. However, there is a useful consistency condition between $\mu'$ and $\mu(\rho_0)$.
Let $f_{\rm cr}(\rho_0)$ denote the free energy density for the fully minimized crystal 
with vacancy concentration $n_{\rm vac,0}$.
Then
\bea
 \label{eq:mucond}
  \mu = \frac{d f_{\rm cr}}{d\rho_0} \stackrel{!}{=} \left. \mu'(\rho_0, n_{\rm vac})
  \right|_{n_{\rm vac}=n_{\rm vac,0}}\;
\eea
This can be shown as follows. Let $\req(\rho_0, n_{\rm vac}; \vect r)$ be the minimizing density profile
for a crystal with fixed bulk density $\rho_0$ and vacancy concentration $n_{\rm vac}$. Using
the expansion in reciprocal lattice vectors (\ref{eq:rhoexpansion}), 
$\rho = \rho_0 + \sum_j \rho_j \exp({\rm i} \vect K_j \cdot \vect r)$, it is seen that the
constrained minimization of Eq.~(\ref{eq:fconstr})  yields
\bea
  \mu'(\rho_0,n_{\rm vac})& = & \frac{1}{a^3(\rho_0,n_{\rm vac})} \int_{\rm cell} d^3r 
   \, \frac{\partial (f^{\rm id}[\req]+f^{\rm ex}[\req])}{\partial \rho_0} \nonumber \\
 \label{eq:muprime}
    & = & \frac{1}{a^3(\rho_0,n_{\rm vac})} \int_{\rm cell} d^3r\, 
   (\ln (\req\Lambda^3) + \mu^{\rm ex}[\req]) \;.
\eea 
Here, the last line follows since 
$\int d^3r \, \partial f^{\rm ex}/ \partial \rho_0 = \int d^3r \, (\delta {\cal F}^{\rm ex}/ 
\delta \rho)\, (\partial \rho/\partial \rho_0)$ and $\partial \rho/\partial \rho_0=1$.
On the other hand, the chemical potential from the crystal equation of state becomes:
\bea
 \mu= \frac{d f_{\rm cr}}{d\rho_0} &=& \left. \frac{\partial f_{\rm cr}}{\partial n_{\rm vac}}\;
  \frac{\partial n_{\rm vac}}{\partial \rho_0}\right|_{n_{\rm vac}=n_{\rm vac,0}} 
   + \frac{1}{a^3(\rho_0,n_{\rm vac,0})}
    \int_{\rm cell} d^3r \, \frac{\partial (f^{\rm id}[\rho_{\rm cr}]+
   f^{\rm ex}[\rho_{\rm cr}])}{\partial \rho_0}  \; 
  \nonumber  \\[2mm]
  &=&  \left. \rho_0 \frac{\partial (F_{\rm cr}/N)}{\partial n_{\rm vac}}\;
  \frac{\partial n_{\rm vac}}{\partial \rho_0}\right|_{n_{\rm vac}=n_{\rm vac,0}}  +
   \mu'(\rho_0,n_{\rm vac,0}) = \mu'(\rho_0,n_{\rm vac,0})
\eea
Thus we see that $\mu(\rho_0)=\mu'(\rho_0,n_{\rm vac,0})$, since the crystal free energy particle per particle
($F_{\rm cr}/N$) is minimal at $n_{\rm vac}=n_{\rm vac,0}$.

\subsection{Basic considerations on single defects}

\label{subsec:defects}

As we have seen in the previous considerations, the appearance of defects enters the
equilibrium density profile in a crystal through the average occupation of a lattice site.
The dominating type of defect in the equilibrium hard sphere crystal are monovacancies
whose properties have been studied before in simulations explicitly \cite{Ben71,Pro01,Kwa08}.
In order to derive a general formula for the constrained chemical potential
$\mu'(\rho_0,n_{\rm vac,0})$ it is useful to discuss the thermodynamics of a crystal containing 
vacancies more in detail. 

Here we follow Ref.~\cite{Pro01} in the subsequent reasoning.
We introduce a system with $M$ lattice sites which contains $n$ monovacancies
at given positions and index thermodynamic quantities with these numbers, such that
e.g. $V_{M,1}$ denotes the volume of a lattice with $M$ sites, 1 fixed vacancy and therefore
$M-1$ particles. Furthermore it is convenient to define by $-f_{\rm vac}$ the change in free energy due to the creation
of a single vacancy at a specific lattice point while keeping the volume and the {number} 
of lattice sites constant:
\bea
  -f_{\rm vac} &= &F_{M+1,1}(M,V_{M+1,0},T)-F_{M+1,0}(M+1,V_{M+1,0},T) \nonumber \\
       &=& - \ln(\rho_0 \Lambda^3)-f_{\rm vac}^{\rm ex} \;. \label{eq:f1def}
\eea 
As usual, the free energy $F(N,V,T)$ is a function of particle number $N$, volume and temperature.
In the second line we have separated $-f_{\rm vac}$ into  the ideal gas contribution
and the excess part $-f_{\rm vac}^{\rm ex}$. 
 Assuming 
no interaction between pairs of monovacancies, the 
total free energy $F_{M,n}$ is 
\bea
  F_{M,n} = F_{M,0} - n f_{\rm vac} =  M f_0 - n f_{\rm vac}
\eea 
where $f_0$ is the free energy per particle (or per lattice site) in a defect--free crystal 
(i.e. precisely the value
of $F/N$ determined in our simulations). 
In order to calculate the equilibrium concentration of vacancies, it is more convenient to switch to
the Gibbs free energy $G_{M,n}(M-n,p,T)$ in a system of $M-n$ particles at constant pressure $p$ and 
temperature $T$. We define $g_{\rm vac}$ as the change in $G$ due to the creation
of a single vacancy at a {\em specific} lattice point:
\bea
   g_{\rm vac} &=& G_{M+1,1}(M,p,T) - G_{M,0}(M,p,T) \nonumber \\
       &=& F_{M+1,1}(M,V_{M+1,1},T) - F_{M,0}(M,V_{M,0},T) + p(V_{M+1,1}-V_{M,0}) \;.
\eea
Using Eq.~(\ref{eq:f1def}) and furthermore
$f_0=F_{M+1,0}(M+1,V_{M+1,0},T)-F_{M,0}(M,V_{M,0},T)$ and $\mu_0=f_0+pV_{M,0}/M$ we find
\bea
  g_{\rm vac} = \mu_0 - f_{\rm vac} \;.
\eea
The total Gibbs free energy $G^{\rm tot}_{M,n}$ includes the entropic contribution due to
the distribution of $n$ vacancies over $M$ lattice sites ($n \ll M$):
\bea
  G^{\rm tot}_{M,n} \approx G_{M-n,0} + n g_{\rm vac}  + n\kt\left( \ln\frac{n}{M} - 1 \right)
\eea  
Minimizing with respect to $n$ yields the
equilibrium concentration of monovacancies $n_{\rm vac,0}$:
\bea
 \label{eq:nvac}
 n_{\rm vac,0} = \frac{n}{M} = \exp(-\beta g_{\rm vac})= \exp(-\beta(\mu_0 - f_{\rm vac})) \;.
\eea

Let us now define an excess chemical potential $\mu'_{\rm Wi}(n_{\rm vac})$ for a constrained crystal at a fixed 
vacancy concentration $n_{\rm vac}$ through the free energy of particle insertion (Widom's 
trick). Its excess part can be estimated by the probability $P_{acc}(V_{\rm WS})$ of inserting a particle into
the Wigner--Seitz cell (with volume $V_{\rm WS}$) around the vacancy position  and the probability 
$n_{\rm vac}$ of picking the
vacancy lattice site among all lattice sites. Thus:
\bea
  \mu_{\rm Wi}'(n_{\rm vac}) &  \approx& \ln(\rho_0\Lambda^3) - \kt \ln \left(P_{\rm acc}(V_{\rm WS})\right) 
    - \kt \ln n_{\rm vac} \nonumber \\
                    &=&  f_{\rm vac} - \kt \ln n_{\rm vac} \label{eq:munvacdef}
\eea
The second line follows since
$- \kt \ln P_{\rm acc}(V_{\rm WS})$ is precisely the excess free energy cost $f_{\rm vac}^{\rm ex} $
of {\em removal} of one vacancy \cite{Pro01}. We see immediately that in equilibrium, 
$ n_{\rm vac} = n_{\rm vac,0}$, we have $\mu_{\rm Wi}'(n_{\rm vac,0}) \approx \mu_0$ 
which demonstrates the consistency between
the thermodynamic and insertion route {\em in equilibrium}.  (Note that the
correction to the equilibrium chemical potential {is} only linear in $n_{\rm vac,0}$ \cite{Pro01}.)    
However, for the constrained system the insertion route predicts that $\mu'(\rho_0,n_{\rm vac})$
{\em diverges} upon $n_{\rm vac} \to 0$.   

The system with the constraint of fixed $n_{\rm vac}$ corresponds to the free energy functional
in Eq.~(\ref{eq:fconstr}) and thus we may identify 
$\mu_{\rm Wi}'(n_{\rm vac}) \equiv \mu'(\rho_0,n_{\rm vac})$. {Therefore} the logarithmic increase
of the chemical potential with $n_{\rm vac}\to 0$ is a {stringent test} for fully minimized 
density functional models.
However, we want to point out that physically the divergence of $\mu'$ with 
vanishing vacancy density is not {entirely correct} {as outlined in the following.}
Even in a perfect lattice it is possible to insert another interstitial particle. 
Similarly to $-f_{\rm vac}$ one can define the change in free energy $f_{\rm in}$ due to the creation
of a single interstitial at a specific lattice point while keeping the volume and 
the {number} of lattice sites
constant:
\bea
  f_{\rm in} &= &F_{M,1}(M+1,V_{M,1},T)-F_{M,0}(M,V_{M,0},T)\;.
\eea 
The second index for $F$ and $V$ refers to the number of interstitial particles in the system. 
Therefore it follows that for vanishing vacancy concentration the constrained chemical potential
is given by
\bea
 \mu'(\rho_0,n_{\rm vac} \ll n_{\rm vac,0}) =  f_{\rm in} + {\cal O}(n_{\rm vac})\;.
\eea
Simulation results for the free energies $f_{\rm vac}$ and $f_{\rm in}$ in hard sphere crystals near
coexistence give approximately the magnitudes 8 $\kt$ and 34 $\kt$, respectively \cite{Pro01,Kwa08}.   
Since $f_{\rm in} \gg f_{\rm vac}$, it is clear that the constrained chemical potential should
exhibit the logarithmic divergence upon $n_{\rm vac} \to 0$ down to very small vacancy concentrations.  
(At coexistence, $\mu'(\rho_0,n_{\rm vac,0}) = \mu_0 \approx 16$ $\kt$. For smaller $n_{\rm vac}$, $\mu'$ should rise
up to approximately $f_{\rm in} \approx 34$ $\kt$ and then level off.) 

\subsection{Previous results in FMT}

A fully three--dimensional minimization of FMT aiming at the crystal profile 
has not been carried out before. In Tarazona's ground--breaking work \cite{Tar00}
the density profile was parametrized as
\bea
  \rho_{\rm cr} (\vect r) &= & \sum_{{\rm lattice\;sites}\;i} (1-n_{\rm vac})\,\left(\frac{\alpha}{\pi}\right) ^{\frac{3}{2}}
  \,\exp\left( -\alpha (\vect r - \vect r_i)^2\right) \,\left(1+ K_4 \alpha^2 f_4(\vect r- \vect r_i) 
      \right) \;, \\
   && f_4 (\vect r = (x,y,z)) = x^4+y^4+z^4- \frac{3}{5} r^4 \;.
\eea    
Here, $f_4$ is the leading term for the unit cell
anisotropy in cubic lattices.  The free energy per particle was minimized with respect
to $n_{\rm vac}$, $\alpha$ and $K_4$ using the
Rosenfeld tensor functional (\ref{eq:phi_hs}) and (\ref{eq:frf}). The
anisotropy turned out to be unimportant for the values of $F_{\rm cr}/N$ (modifying it by less
than $10^{-3}$ $\kt$). Both $F_{\rm cr}/N$ and $\alpha$ were shown to be in good agreement 
with the old simulation data of Ref.~\cite{You74}. No clear free energy minimum was found
for a nonzero $n_{\rm vac}$, indicating that $n_{\rm vac,0}<10^{-8}$.  

Concerning the issue of the equilibrium vacancy concentration $n_{\rm vac,0}$ in FMT, there
are two more, partially contradictory statements in the literature. In Ref.~\cite{Ros97}
it was argued that the correct 0d limit of a density functional (for particles strictly localized
to their lattice sites) should always lead to a finite, but small $n_{\rm vac,0}$.
The 0d excess free energy is given by $\beta F^{\rm ex}_{\rm 0d} = \eta +(1-\eta)\,\ln (1-\eta)$ 
with a corresponding excess chemical potential $\beta \mu^{\rm ex}_{\rm 0d} = -\ln(1-\eta)$. Since
the packing fraction at each lattice site is corresponds to $1-n_{\rm vac}$,
one finds $\beta \mu^{\rm ex}_{\rm 0d} = -\ln n_{\rm vac}$. The equilibrium vacancy concentration
follows upon identification of $\mu^{\rm ex}_{\rm 0d}(n_{\rm vac,0})$ with the crystal chemical potential
$\mu_0$ as $n_{\rm vac,0} = \exp(-\beta\mu^{\rm ex}_{\rm 0d})$. According to this
argument one would expect an
equilibrium vacancy concentration $n_{\rm vac,0} \sim 10^{-8}$ at coexistence.
We observe that the divergence of $\beta \mu^{\rm ex}_{\rm 0d}$ 
is precisely of the type derived before for the constrained chemical potential
$\mu'(\rho_0,n_{\rm vac})$ (see Eq.~(\ref{eq:munvacdef})). 
However, the plain identification $\mu^{\rm ex}_{\rm 0d} \equiv \mu'$ is incorrect due to the
neglect of the free energy of vacancy formation. This explains the four orders of magnitude 
difference in $n_{\rm vac,0}$ when compared with simulations \cite{Ben71, Kwa08}.
Another approach was taken in Ref.~\cite{Gro00} to calculate 
$n_{\rm vac,0}$. There, the Rosenfeld functional (among others) was minimized in a perturbative approach
assuming isotropic density distributions around lattice sites and an expansion around the close--packing 
limit. A free energy minimum was found for values of $n_{\rm vac,0}$ consistent with simulation results.
However, we will demonstrate below that this finding is not consistent with our full minimizations. 

The success of the density parametrization using isotropic Gaussians and zero vacancy concentration
inspired the works of Ref.~\cite{Lut06} to investigate non-fcc crystals and of 
Refs.~\cite{Son06,Son08} to treat binary systems and the crystal--fluid interface within FMT.
In the latter work, the interface density profile was parametrized in an intuitive way, however,
in this way one cannot ensure that crystal and fluid are in chemical equilibrium (see Sec.~\ref{sec:results} below).

\subsection{Numerical solution of the FMT Euler--Lagrange equation}

In {actual} calculations, we determine the constrained crystal profile
$\req(\rho_0, n_{\rm vac}; \vect r)$ by a full minimization in three--dimensional real space.
For such a three-dimensional problem, the density profile $\rho$
and 11 weighted densities (two scalar densities $n_2,\, n_3$, three vector densities $(\vect n_2)_i$
for $i=\{x,y,z\}$ and six tensor densities $(n_t)_{ij}$ for $\{ij\}=\{xx,yy,zz,xy,xz,yz\}$)
need to be discretized on a three--dimensional grid covering the cubic unit cell.  
Usually we chose grids with dimensions 64$^3$ (for lower densities around the
coexistence density $\rho_{\rm coex}\sigma^3\approx 1.04$) up to 256$^3$ (for higher densities). 
{Here, $\sigma=2R$ is the hard sphere diameter.}
The necessary convolutions were computed using Fast Fourier Transforms. 
With prescribed $\rho_0$ and $n_{\rm vac}$, the constrained functional (\ref{eq:fconstr}) is minimized 
through Picard iteration (with mixing) of the Euler--Lagrange equation (\ref{eq:rhoeq}). 
A new profile $\rho_{i+1}$ is determined from an old profile $\rho_i$ and an appropriate
$\mu'_i$ through
\bea
   \rho_{i+1} & = & \alpha\, \rho'_{i+1} + (1-\alpha)\,\rho_i \;, \\
   \rho'_{i+1} &=& \exp\left( -\beta\frac{\delta {\cal F^{\rm ex}}}{\delta \rho(\vect r)}[\rho_i]
  + \beta \mu'_i  \right) \;.
\eea
Here, $\mu'_i$ is determined such that $\int_{\rm cell} d^3r\, \rho_{i+1}=4(1-n_{\rm vac})$. 
The mixing parameter $\alpha$ is of the order of 0.01. The iteration was stopped when
the relative deviation between $\mu'_i$ and $\mu'(\rho_0,n_{\rm vac})$ from Eq.~(\ref{eq:muprime})
was below $5\cdot 10^{-6}$. The iteration procedure was stabilized by two means: (i) enforcing
the physical requirement $n_3(\vect r) \le 1-n_{\rm vac}$ at each iteration step since the
singularity at $n_3=1$ (see the functional in Eq.~(\ref{eq:phi_hs})) is avoided in that manner. 
(ii) enforcing the point symmetry of the fcc crystal in the density profile in each iteration step.
In each iteration, this point symmetry is slightly violated by numerical inaccuracies. Without correction
and using $\alpha \sim 0.01$,
this symmetry violation quickly grows and eventually leads to numerical singularities.
Only for $\alpha \alt 10^{-5}$, convergence was achieved without explicit enforcement of the
point symmetry at the price of an increase in computation time by a factor 100--1000.     
For a given $\rho_0$, $n_{\rm vac}$ is varied and the location of the minimum is 
checked using the consistency condition (\ref{eq:mucond}). However, for $n_{\rm vac}<10^{-5}$
it proved to be hard to arrive at a convergent solution.

\section{Monte Carlo simulations}

\label{sec:mc}

\subsection{Computation of absolute {free} {energies}}

The computation of absolute free energies poses a problem to Monte Carlo
simulation, because it requires the evaluation of the partition function. 
For most systems that have an infinite and continuous state space, the 
partition function cannot be computed {directly}. However, one can compute free 
energy differences and derivatives by MC simulation. Hence, if 
there is a suitable reference 
system, the free energy of which is known analytically, free energies
can be extracted from MC simulation. Here we use a technique that was 
recently introduced by Schilling and 
Schmid \cite{SchillingSchmid09}. The technique {extends the well-established} thermodynamic 
integration with respect to the harmonic crystal (Einstein {crystal}) \cite{FrenkelLadd84} to 
disordered reference states and tether potentials that are not harmonic. 
We compare our results to DFT and to simulation results obtained by Vega and Noya using the Einstein 
Molecule (EM) technique (a variant of the Einstein {crystal} that avoids having to 
correct for the center of mass motion of the system \cite{VegaNoya07}).

{We used systems of size $N=1728$ and potential
wells of radius $r_{\rm cutoff}=0.75\;\sigma$. The path of the
thermodynamic integration for each density was subdivided into
simulations of 35 different values of the coupling strength $\epsilon$
between 0 and 80, each of which consisted of $10^4$ equilibration
sweeps and $10^6$ sweeps of averaging (where one sweep consisted of $N$ 
attempted particle moves). There are two sources of error: The first 
one results from the prodecure of integration and can
be estimated to $(\beta\Delta F)_{\rm int}/N=0.001$. The {second is} statistical{. These} errors
were computated by using the Jackknife algorithm with 1000 subsets on
each part of every integration. With this the statistical errors equally
amount to $(\beta\Delta F)_{\rm stat}/N=0.001$.}

\begin{table}
  \caption{\label{tab:Comp}Comparison of free energies (a) calculated
    using the algorithm of \cite{SchillingSchmid09} and (b) the
    results obtained with the EM method from \cite{VegaNoya07}. For the DFT results, the
  White Bear II functional and $n_{\rm vac}=10^{-4}$ was used. 'Gauss' refers to minimization
  using the Gaussian approximation (Eq.~(\ref{eq:Gauss_ansatz})).
  The free energies according to the Speedy equation of state have been determined using
  Eq.~(\ref{eq:fspeedy}).
 In order to obtain numbers, 
  $\Lambda=\sigma$ has been used.
 }
  \begin{ruledtabular}
    \begin{tabular}{lllllll}
      $\rho_0\sigma^3$ & $\beta F/N^{(a)}$  &
      $\beta F/N^{(b)}$  & $\beta F/N^{(b)}$  & $\beta F_{\rm DFT}/N$  & $\beta F_{\rm DFT}/N$ & $\beta F_{\rm Speedy}/N$   \\
        & ($N=1728$) & ($N=2048$) & ($N\to \infty$) & (Gauss) & (full min.) &  \\      
      \hline
       1.00     &  4.530(2)  &            &           & 4.541 & 4.539 & 4.532 \\
       1.04086  &  4.960(2)  &  4.955(1)  & 4.9590(2) & 4.979 & 4.977 & 4.961 \\
       1.049    &  5.048(2)  &            &           & 5.069 & 5.067 & 5.049 \\
       1.08     &  5.397(2) &            &           & 5.424 & 5.422 & 5.398 \\
       1.09975  &  5.631(2)  &  5.627(1)  & 5.631(1)  & 5.660 & 5.658 & 5.631 \\
       1.11     &  5.756(2) &            &           & 5.787 & 5.785 & 5.756 \\
       1.14     &  6.142(2) &            &           & 6.174 & 6.172 & 6.140 \\
       1.15000  &  6.277(2)  &  6.269(1)  & 6.273(2)  & 6.310 & 6.308 & 6.275 \\
    \end{tabular}
  \end{ruledtabular}
\end{table}

\subsection{Density Distribution}

A {second set of} simulations has been carried out to sample the density distribution of the hard sphere crystal
unit cell with high accuracy.   A perfect {fcc} crystal has been set up in a cubic box with 
fixed side lengths, periodic boundary conditions and a particle number of $N = 4 \cdot n^3$ Particles, 
which corresponds to $n$ unit cells along one side of the box.

 The simulations have been run on a standard octocore CPU. 
The results for the density distributions presented below are based on simulations of a system 
consisting of $N_{n=13} = 8788$ particles.
Different system sizes of $N_{n=9}= 2916$, $N_{n=11}= 5324$ or $N_{n=15}= 13500$ particles 
have been used to study the extrapolation of the average Gaussian width $\alpha$ 
(see Eq.~(\ref{eq:Gauss_ansatz})) for $N \rightarrow \infty$.  Snapshots of the system 
configuration have been taken every 30 sweeps. After each sweep, appropriate global shifts of particle coordinates were
applied to keep the center of mass fixed.   
The simulated bulk densities vary from $\rho_0 \sigma^3 = 1.04$ to 
$\rho_0 \sigma^3= 1.30$, with the corresponding density distributions averaged over
$(1\dots 2) \cdot 10^{11}$ snaps{hots} (depending on exact acceptance rate). This corresponds to 
approximately 18 days of CPU time for a single density distribution.  
Error estimates for $\rho_0\sigma^3 = 1.04$ and $\rho_0\sigma^3 = 1.20$
have been obtained using the Jackknife algorithm, with the unit cell histograms {divided} into 1000 subsets. 
All simulations have been started with 100,000 sweeps of equilibration. In order to obtain the density
distributions in a single unit cell,  the system snapshots of all unit cells were mapped onto
one unit cell providing us with a 3D histogram with a resolution of 80 bins per unit cell length.


\section{Results}

\label{sec:results}

\subsection{Free Energies and coexistence densities}

In order to connect to the previous simulation work in Ref.~\cite{VegaNoya07}, 
we studied hard sphere systems with particle
densities $\rho_0\sigma^3=1.04086$, $1.09975$ and $1.15$. Additionally,
we also considered more density points and
the values for the free energy per particle  $F/N$ for all our calculations are reported in
Tab.~\ref{tab:Comp}.
All simulation results were obtained with $n_{\rm vac}=0$,
whereas in the DFT results (using the {White} Bear II (WBII) functional) 
$n_{\rm vac}=10^{-4}$ was chosen which did not affect the values for $F/N$
to the accuracy shown. Our simulation results and the results from Ref.~\cite{VegaNoya07}
are consistent with each other on the level of 0.05 \%. The DFT results are systematically 
larger than the simulation results, the discrepancy here is also not larger than 0.5\%. 

The last column in Tab.~\ref{tab:Comp} gives the corresponding free energy results as derived
from the popular equation of state proposed by Speedy \cite{Spe98}. This equation of state is given
in the form
\bea
  \frac{\beta p_{\rm Speedy}}{\rho} &=& \frac{\beta p_{\rm sing}}{\rho} + 3 -
  c_1\; \frac{\frac{\di \rho}{\di \rho_{\rm cp}}-c_2}{\frac{\di \rho}{\di \rho_{\rm cp}}-c_3}\;, \\
   \frac{\beta p_{\rm sing}}{\rho} &=& \frac{3}{\frac{\di \rho_{\rm cp}}{\di \rho}-1} \;,
\eea
where $\rho_{\rm cp}\sigma^3=\sqrt{2}$ is the close--packing density, and $c_1=0.5914$,
$c_2=0.7079$ and $c_3=0.6022$ are
fitting constants determined recently from a fit to a substantial set of pressure data \cite{Alm09}.
The pressure $p_{\rm Speedy}$ exhibits the divergent behavior ($p_{\rm sing}$) for
$\rho\to\rho_{\rm cp}$ as predict by free--volume considerations. In order to obtain the
free energy, thermodynamic integration 
can be applied after the divergent piece has been subtracted and integrated separately:
\bea
 \label{eq:fspeedy}
  \frac{\beta F_{\rm Speedy}}{N}(\rho) &=& \int_{\rho_{\rm cp}}^{\rho} 
   \frac{\beta (p_{\rm Speedy}-p_{\rm sing})}{\rho^2}  - 3 \ln[(\rho_{\rm cp}-\rho)\sigma^3] + C\;.
\eea
Here, $C$ is an integration constant which has been quoted in the literature \cite{Gro00b}
from a fairly old simulation \cite{Ald68}
as $C = 2.843 \pm 0.040$. Fitting $C$ by using Eq.~(\ref{eq:fspeedy}) to our MC free energy data
we find the improved estimate $C = 2.8247 \pm 0.0006$. 


For DFT, we used the Gaussian approximation to the density profiles to calculate the thermodynamic
properties of liquid--solid coexistence via the Maxwell construction. The result is given in 
Tab.~\ref{tab:coex} for the three investigated fundamental measure models and compared to
very recent simulation results \cite{Zyk10}. For the tensor modified Rosenfeld (RF) and White Bear (WB) functionals we
recover the results quoted in Refs.~\cite{Tar00,Rot02}. For the RF functional, the coexistence densities
are substantially smaller than the corresponding densities for WB and WBII. This is entirely due to      
the insufficient accuracy of the Percus--Yevick equation of state on the fluid side which underlies
the RF functional. The WB and WBII functionals reduce to the Carnahan--Starling equation of state for
homogeneous densities and thus their thermodynamic description of liquid--solid coexistence is
satisfactory. 

\begin{table}
 \caption{\label{tab:coex} Coexisting fluid ($\rho_{\rm fl}$) and crystal ($\rho_{\rm cr}$)
  densities (the corresponding packing fractions are given in brackets), as well as the
  chemical potential $\mu_{\rm coex}$ and the pressure $p_{\rm coex}$ at coexistence for the
  three investigated DFT models. Here, RF is the tensor modified Rosenfeld functional 
  with the free energy density determined by Eqs.~(\ref{eq:phi_hs}) and (\ref{eq:frf}),
  WB is the tensor modified White Bear functional (Eqs.~(\ref{eq:phi_hs}) and (\ref{eq:fwb})), and
  WBII is the tensor modified White Bear II functional (Eqs.~(\ref{eq:phi_hs}) and (\ref{eq:fwbII})). 
  The MC results are taken from Ref.~\cite{Zyk10}.
  In order to obtain numbers, $\Lambda=\sigma$ has been used.}
 \begin{ruledtabular}
  \begin{tabular}{lllll}
       & $\rho_{\rm fl}\sigma^3\;(\eta_{\rm fl})$ & $\rho_{\rm cr}\sigma^3\;(\eta_{\rm cr})$ &
   $\beta \mu_{\rm coex}$ & $\beta p_{\rm coex} \sigma^3$ \\  \hline
    RF   & 0.892$\;$(0.467)   & 0.984$\;$(0.515) & 14.42 & \phantom{1}9.92 \\
    WB   & 0.934$\;$(0.489)   & 1.022$\;$(0.535) & 15.75 & 11.28 \\
    WBII & 0.945$\;$(0.495)   & 1.040$\;$(0.544) & 16.40 & 11.89 \\ \hline
    MC   & 0.940$\;$(0.492)   & 1.041$\;$(0.545) &       & 11.576 \\
  \end{tabular}
 \end{ruledtabular}
\end{table}

\subsection{ Vacancy concentration and constrained chemical potential}

\begin{figure}
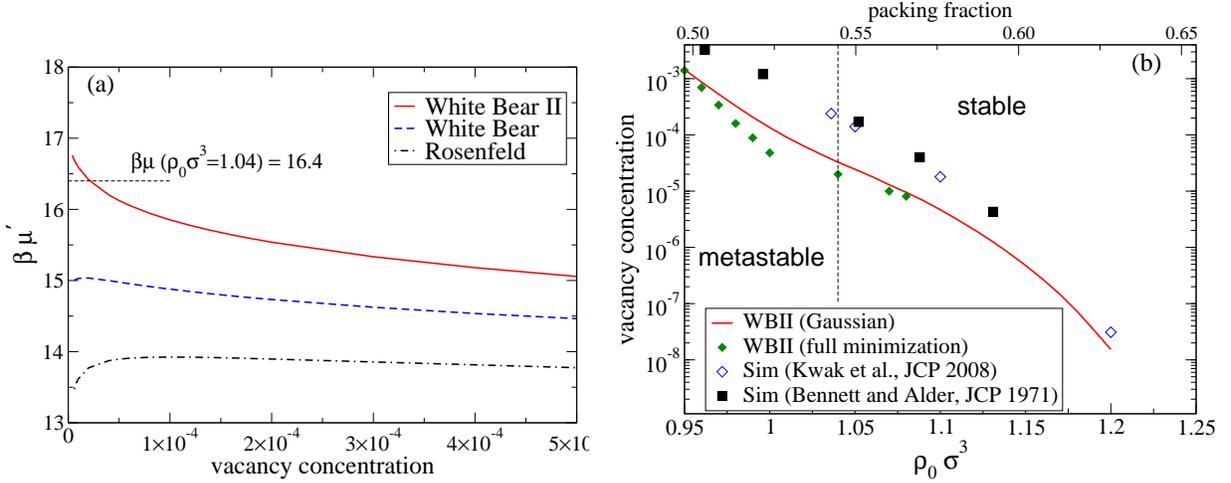

 \begin{center}
   \epsfig{file=fig2a.eps, width=8cm}
   \epsfig{file=fig2b.eps, width=8cm}
 \end{center}
  \caption{\label{fig:vac}
  (a): The constrained chemical potential $\mu'(\rho_0,n_{\rm vac})$ as obtained by full minimization 
  of the three DFT models at the bulk density $\rho_0\sigma^3=1.04$. 
  The dashed line shows the value for the chemical potential following from the 
  thermodynamic definition of $\mu=d f_{\rm cr}/d\rho_0$
  where $f_{\rm cr}$ is the free energy density. It is equal for the three DFT models to the given accuracy.    
  If and only if $\mu'(\rho_0,n_{\rm vac})=\mu$, the free energy per particle is minimal and thus thermodynamic
  consistency holds. (b): Equilibrium vacancy concentration vs. bulk density as obtained for the
  WBII functional (full line--Gaussian approximation, filled diamonds--full minimization) and compared
  to available simulation results (open diamonds--Ref.~\cite{Kwa08}, filled squares--Ref.~\cite{Ben71}).}
\end{figure}

In Sec.~\ref{subsec:defects} we have derived an expression for the constrained chemical potential
$\mu'(\rho_0,n_{\rm vac})$ (see Eq.~(\ref{eq:munvacdef}){)}.  
Furthermore we recall that $\mu'(\rho_0,n_{\rm vac})$ is precisely the Lagrange multiplier in the
constrained minimization of the unit cell free energy, see Eq.~(\ref{eq:fconstr}). We have examined its
dependence on $n_{\rm vac}$ and {the} bulk density $\rho$ for the three functionals with the surprising result
that only in the case of the White Bear II functional $\mu'(\rho_0,n_{\rm vac})$ shows a weakly divergent behavior
as $n_{\rm vac} \to 0$. The divergence appears to be weaker than $-\ln n_{\rm vac}$, however.  
Furthermore only for the White Bear II functional the consistency condition 
(\ref{eq:mucond}) is fulfilled for a (small and) finite equilibrium vacancy concentration $n_{\rm vac,0}$.   
There is no minimum for the free energy per particle $F/N$ upon variation of $n_{\rm vac}$ for the cases
of the Rosenfeld and the White Bear functional (neither in the Gaussian approximation nor  for 
full minimization). This is consistent with Tarazona's finding of no minimum for $n_{\rm vac}>10^{-8}$ using
the Gaussian approximation in the Rosenfeld functional \cite{Tar00}. As an exemplary result, we show
$\mu'(\rho_0,n_{\rm vac})$ for $\rho_0 \sigma^3=1.04$ (coexistence) for the three functionals, see 
Fig.~\ref{fig:vac} (a). The large discrepancies between results for the three functionals is 
somewhat surprising, given the fact that $F/N$ varies only very little (in the Gaussian approximation
we have $\beta F/N=4.929$ [RF], 4.912 [WB], 4.970 [WBII] at this density). Our results for the explicit
minimization also question the reliability of the approach taken in Ref.~\cite{Gro00} to calculate 
$n_{\rm vac,0}$. 
It appears that the equilibrium vacancy concentrations and corresponding free energy minima are
artefacts of the approximations used therein (isotropic density distributions around lattice sites and an expansion around the close--packing limit).
  
In Fig.~\ref{fig:vac} (b) we show the variation of the equilibrium vacancy concentration with
the bulk density for the WBII functional. There is reasonable agreement between the Gaussian approximation and
the full minimization. However, the predicted $n_{\rm vac,0}$ is consistently smaller (up to one order of magnitude)
than available simulation results. Nevertheless one should keep in mind that the DFT results do not follow
from an explicit computation of the free energy of a vacancy $f_{\rm vac}$ (see Eq.~(\ref{eq:f1def})) as the
simulations do. It would be interesting in the future to calculate $f_{\rm vac}$ through an explicit
minimization of DFT around a fixed vacancy. Note that an initial attempt in that direction has been 
undertaken in Ref.~\cite{Sin05} using the MWDA. 

An important implication arises from the fact that the consistency condition (\ref{eq:mucond}), 
$\mu'(\rho_0,n_{\rm vac,0})=d f_{\rm cr}/d \rho_0$, can be fulfilled only for the WBII functional. 
It means that a free DFT minimization
of the fluid--crystal interface which is consistent with the coexistence data from the Maxwell construction 
(see Tab.~\ref{tab:coex}) 
will not be possible with the WB and the RF functionals. 
This follows since the fluid chemical potential at coexistence
does not match the crystal chemical potential  obtained by full mimimization.

\subsection{Density distributions}

The density distribution in the hard sphere crystal consists of nearly isolated density peaks
around the lattice sites,
\bea
   \rho_{\rm cr}(\vect r) = \sum_{{\rm lattice\;sites}\;i} \rho(\vect r - \vect r_i)
\eea
with no appreciable overlap in the tails of $\rho(\vect r)$.
In first approximation, $\rho(\vect r)$ is a Gaussian with a width parameter $\alpha$,
\bea
 \label{eq:rhoG}
  \rho(\vect r) \approx \rho_{\rm G}(r) = \left(\frac{\alpha}{\pi}\right)^{\frac{3}{2}}
  \exp(-\alpha r^2)\;.
\eea 
We will analyze the deviations from the Gaussian form in terms of an average radial deviation 
$f_{\Delta {\rm  G}}(r)$  and an anisotropic deviation $f_{\rm aniso}(\vect r)$:
\bea
 \label{eq:faniso_def}
  \rho(\vect r) \approx \rho_{\rm G}(r)\; f_{\Delta {\rm  G}}(r)\; f_{\rm aniso}(\vect r)\;.
\eea 
The average radial deviation will be parametrized as
\bea
  \label{eq:fDG}
  f_{\Delta {\rm  G}}(r) = \exp\left[ b_2\,\alpha r^2 + b_4(\alpha r^2)^2 + b_6(\alpha r^2)^3\right]\;, 
\eea
where $b_2,b_4,b_6 \ll 1$ are expected to be small. For the analysis of the 
directional anisotropy we apply a polynomial expansion in the form:
\bea
 \label{eq:faniso}
  f_{\rm aniso}(\vect r) &=& 1 + K_4\,\alpha^2\, \left(x^4+y^4+z^4-\frac{3}{5}r^4\right) + 
    K_6\,\alpha^3\, \left(x^6+y^6+z^6-\frac{3}{7}r^6\right) \;.
\eea
This corresponds to the leading two terms in the cubic cell asymmetry (consistent with the point symmetry
of the fcc lattice).\footnote{The density distribution around a lattice site
can be expanded as $\rho(\vect r) = \rho_0(r) + \sum_i \rho_i(r) \hat x_i + \sum_{ij} \rho_{ij}(r)
\hat \hat x_i \hat x_j + \dots$, where $\hat x_i= x_i/r$ and the expansion coefficients 
$\rho_{ij\dots}(r)$ are isotropic functions. From symmetry we have $\rho(\vect r) = \rho(-\vect r)$
and $\rho_{ij\dots}(r)=\rho_{P(i)P(j)\dots}(r)$ where $P$ is a permutation of the Cartesian indices.
This implies that all expansion coefficients with an odd number of indices are zero and that
$\rho_{11}=\rho_{22}=\rho_{33}$, giving only an isotropic correction to second order (which can be 
absorbed into $\rho_0(r)$). The lowest nontrivial expansion coefficients are $\rho_{1111}$ and
$\rho_{1122}$ which are not independent of each other since $1=(\hat x_1^2+\hat x_2^2+\hat x_3^2)^2$.
In our fits, we have chosen the radial dependence $\rho_{1111}(r)=K_4\alpha^2\, \rho_{\rm G}(r)\, f_{\Delta {\rm  G}}(r)\,r^4$ and also
demanded that the angular integral of the anisotropy corrections over the unit sphere vanishes.
This leads to an isotropic offset, such that the isotropic piece  for our case becomes
$\rho_0(r) = \rho_{\rm G}(r)\, f_{\Delta {\rm  G}}(r)\,(1 - (3/5)K_4\alpha^2\,r^4)$.}

\subsubsection{Gaussian width parameter}

\begin{figure}
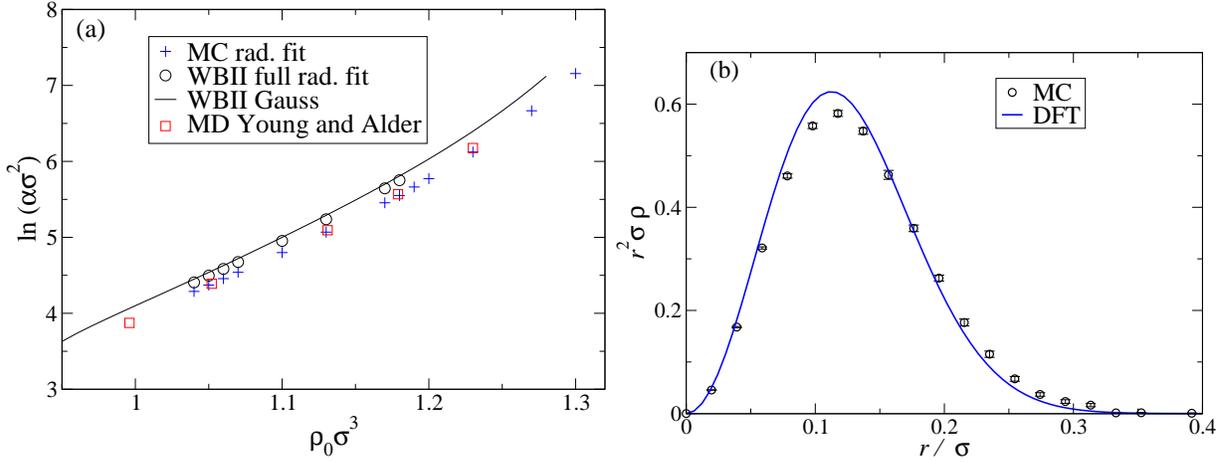

 \begin{center}
   \epsfig{file=fig3a.eps, width=8cm}
   \epsfig{file=fig3b.eps, width=8cm}
 \end{center}
  \caption{\label{fig:alpha}
  (a): Logarithm of the Gaussian width parameter $\alpha$ vs. bulk density $\rho_0$:
  DFT--WBII in Gaussian approximation (full line), DFT--WBII in full minimization (circles), extrapolation
  to the thermodynamic limit in MC ($+$ symbols) and results from Ref.~\cite{You74} (squares).
  (b): radial probability $r^2\rho(r)$ in [100] direction for the bulk density 
  $\rho_0\sigma^3=1.04$. Comparison between DFT and our simulations. 
  }
\end{figure}

For the DFT results, the width parameter $\alpha$ is the only minimization parameter in the 
Gaussian approximation once the normalization is fixed. From the results of the full minimization
we determined $\alpha$ by a global fit with the Gaussian form (\ref{eq:rhoG}) 
to the lattice peak density distribution. The same was done using the MC data, additionally the 
value $\alpha_\infty$  in the thermodynamic limit was determined by the extrapolation from the values at
finite box length $L$ through the relation $\alpha_N= A/N^{1/3} + \alpha_\infty$ \cite{You74}.
For the densities $\rho_0\sigma^3=1.05$ and 1.13 we checked and confirmed this scaling for the 
four values
$N=2916$, 5324, 8788 and 13500. For the other bulk density values, we used $N=5324$ and 13500
to determine $\alpha_\infty$.

In Fig.~\ref{fig:alpha} (a) we compare the DFT results for $\alpha$ with $\alpha_\infty$
from our MC simulations and a corresponding width parameter extracted from the work of
Young and Alder \cite{You74}. There the mean square deviation was determined which we converted
to the Gaussian width parameter by assuming the Gaussian form for $\rho(\vect r)$:
$\alpha= 3/(2\langle r^2\rangle )$. There is excellent agreement between the two simulations
and also fair agreement between DFT and simulations. The Gaussian peaks in DFT are narrower
than the simulated peaks which is similar to (M)WDA results although in (M)WDA the quantitative deviation
is already considerable (compare e.g. with Tab.~I in Ref.~\cite{Tej95}). 
{ The radial probability, proportional to $r^2\rho(r)$, along the $[100]$ direction is shown in 
Fig.~\ref{fig:alpha} (b). As a remark, earlier MC data for the radial probability were 
erroneously scaled in the graphical presentations
of Ref.~\cite{Ohn93}. }

\begin{figure}
 \begin{center}
   \epsfig{file=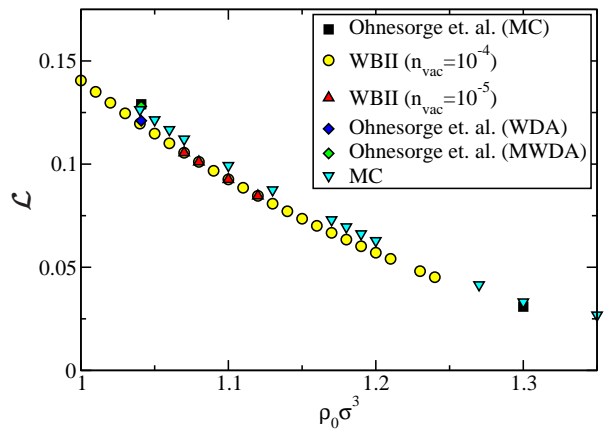, width=8cm} 
 \end{center}
  \caption{\label{fig:lindemann}
	{Lindemann parameter ${\cal L}$ vs. bulk density $\rho_0$ for Monte Carlo simulation (MC) and DFT, also 
	in comparison with Ref.~\cite{Ohn93} (Ohnesorge et. al.). Data for (M)WDA from Ref.~\cite{Ohn93} 
	are taken for full minimization. }
}
\end{figure}

{
In order to quantify the spread of the density distribution
around a solid peak it is convenient to define the Lindemann parameter
\cite{Lin10,Ubb78} as the dimensionless root mean-square displacement:
}
\bea
 \label{eq:lindemann}
	{\cal L} = \frac{1}{r_{nn}} \sqrt{\int_{WSC}d^3 r r^2\rho(\vect r)} \;. 
\eea
{
Here the spatial integration is over a Wigner-Seitz cell ($WSC$) centered 
around a lattice position at the origin and 
$r_{nn}= \sigma \left(\sqrt{2}/\rho_0\right)^{1/3}$ denotes the distance 
between two nearest neighbours in the crystal lattice. Data for the Lindemann 
parameter ${\cal L}$ versus bulk density $\rho_0$ are shown 
in Fig.~\ref{fig:lindemann}. Clearly, $\cal L$ is about $0.13$ at melting and 
decreases with increasing density. The Monte Carlo data published earlier 
in Ref.~\cite{Ohn93} agree with those from our simulations. All density functionals 
considered here (WDA, MWDA, WBII) yield Lindemann parameters which are only slightly 
lower than the simulation data. {However, the density profiles from (M)WDA and 
WBII differ, the agreement in the value of ${\cal L}$ is due to an unphysical high interstitial density 
in the (M)WDA profiles \cite{Ohn93}.} 
}

\subsubsection{Deviations from the Gaussian form and anisotropy}

\begin{figure}
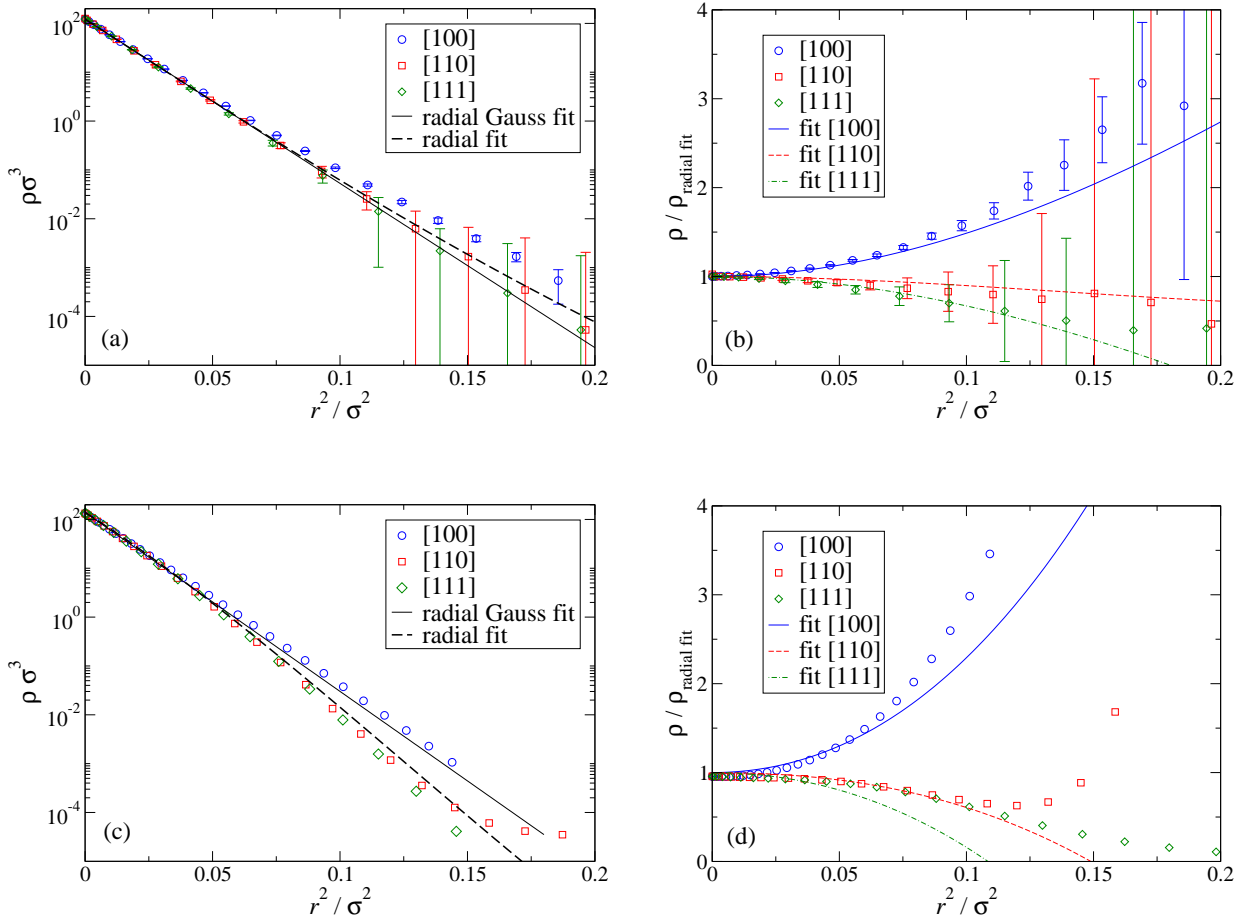

 \begin{center}
   \epsfig{file=fig5a.eps, width=8cm} \hspace{5mm}
   \epsfig{file=fig5b.eps, width=7.6cm} \\[1cm]
   \epsfig{file=fig5c.eps, width=8cm} \hspace{5mm}
   \epsfig{file=fig5d.eps, width=7.6cm}
 \end{center}
  \caption{\label{fig:rho_r1.04}
  Lattice site density distributions along the lattice directions [100], [110] and [111]
  for the bulk density $\rho_0\sigma^3=1.04$. Panels (a) and (b) show MC results ($N=8788$), panels (c) and (d) results
  from DFT--WBII. Panels (a) and (c) show $\rho$ vs. $r^2$ in logarithmic scale, thus illustrating
  the deviation from a Gaussian form (straight line). The full line here is a fit to the Gaussian form
  $\rho_{\rm G}$ (Eq.~(\ref{eq:rhoG})) with the parameter $\alpha= 77.5$ (MC) and $\alpha=84.4$ (DFT--WBII).
  The dashed line is a fit to the non--Gaussian form $\rho_{\rm G}\,f_{\Delta{\rm G}}$ (see Eq.~(\ref{eq:fDG})) 
  with the parameters $b_2=-0.011$, $b_4=0.0021$, $b_6=-0.0002$ (MC) and $b_2=0.090$, $b_4=-0.029$, $b_6=0.0009$
  (DFT--WBII). Panels (b) and (d) show the density along the three lattice directions divided by
   $\rho_{\rm G}\,f_{\Delta{\rm G}}$. The lines show the corresponding anisotropies along the
  three lattice directions resulting from a fit to $f_{\rm aniso}$ (see Eq.~(\ref{eq:faniso})) with the 
  parameter $K_4=0.022$ (MC) and $K_4=0.039$ (DFT--WBII).
}
\end{figure}

\begin{figure}
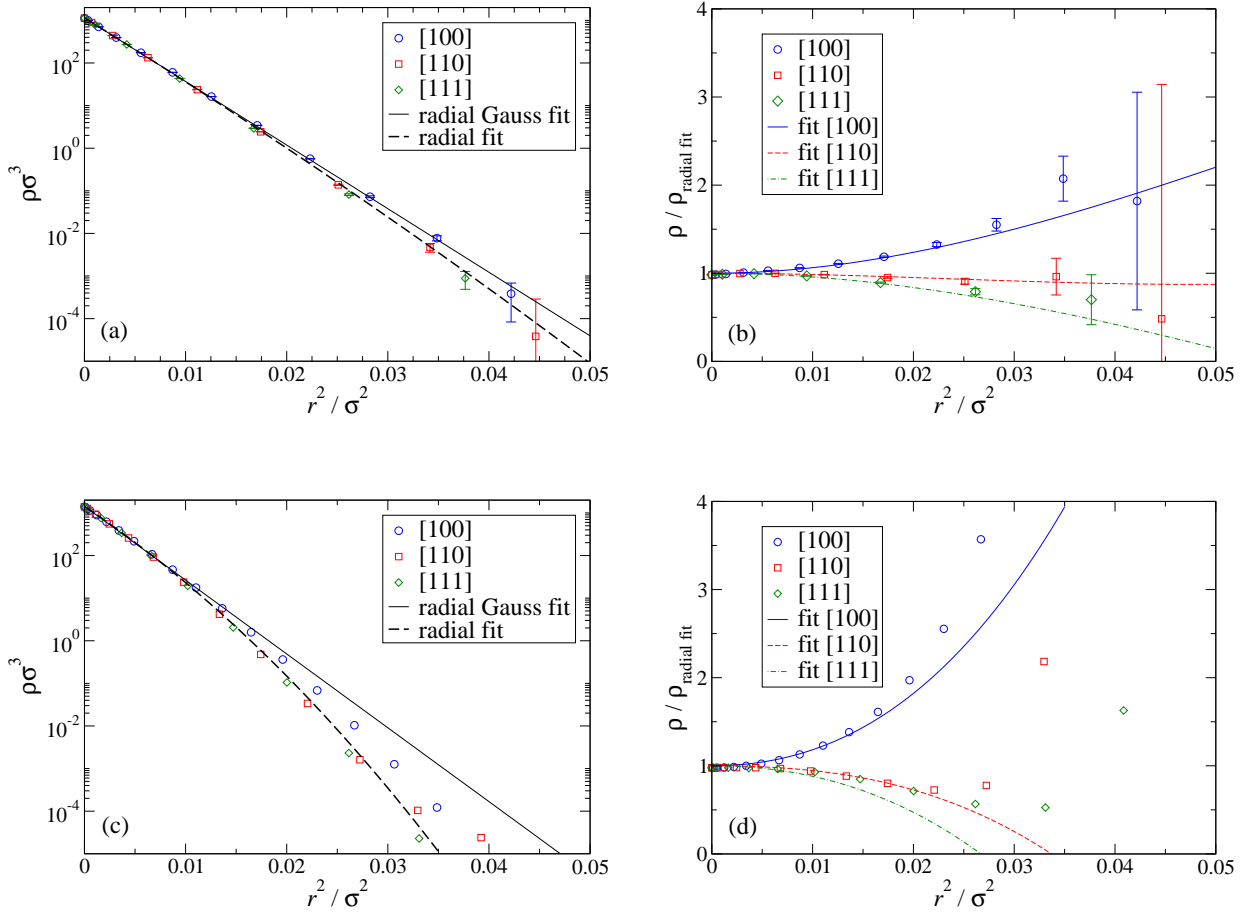

 \begin{center}
   \epsfig{file=fig6a.eps, width=8cm} \hspace{5mm}
   \epsfig{file=fig6b.eps, width=7.6cm} \\[1cm]
   \epsfig{file=fig6c.eps, width=8cm} \hspace{5mm}
   \epsfig{file=fig6d.eps, width=7.6cm}
 \end{center}
  \caption{\label{fig:rho_r1.20}
  Lattice site density distributions along the lattice directions [100], [110] and [111]
  for the bulk density $\rho_0\sigma^3=1.20$. Panels (a) and (b) show MC results ($N=8788$), panels (c) and (d) results
  from DFT--WBII. Panels (a) and (c) show $\rho$ vs. $r^2$ in logarithmic scale, thus illustrating
  the deviation from a Gaussian form (straight line). The full line here is a fit to the Gaussian form
  $\rho_{\rm G}$ (Eq.~(\ref{eq:rhoG})) with the parameter $\alpha= 343.7$ (MC) and $\alpha=399.0$ (DFT--WBII).
  The dashed line is a fit to the non--Gaussian form $\rho_{\rm G}\,f_{\Delta{\rm G}}$ (see Eq.~(\ref{eq:fDG})) 
  with the parameters $b_2=0.014$, $b_4=-0.0054$, $b_6=-0.00002$ (MC) and $b_2=0.075$, $b_4=-0.026$, $b_6=-0.0002$
  (DFT--WBII). Panels (b) and (d) show the density along the three lattice directions divided by
   $\rho_{\rm G}\,f_{\Delta{\rm G}}$. The lines show the corresponding anisotropies along the
  three lattice directions resulting from a fit to $f_{\rm aniso}$ (see Eq.~(\ref{eq:faniso})) with the 
  parameter $K_4=0.014$ (MC) and $K_4=0.025$ (DFT--WBII).
  }
\end{figure}

In Figs.~\ref{fig:rho_r1.04} and \ref{fig:rho_r1.20} we show in an exemplary way the
density distributions in the principal lattice directions [100], [110] and [111]
for the densities $\rho_0\sigma^3=1.04$ (near coexistence) and $\rho_0\sigma^3=1.20$,
respectively. The simulation data are always very close to the Gaussian form 
with the coefficient $b_4$ of the leading deviation from the Gaussian form being small,
$|b_4| \alt 0.01$ (see panel (a) in Figs.~\ref{fig:rho_r1.04} and \ref{fig:rho_r1.20}). 
Interestingly, $b_4$ changes sign at around $\rho_0\sigma^3=1.10$,
indicating that below that density the distribution is wider than a Gaussian {(larger curtosis)} and above
that density the distribution is narrower than a Gaussian {(smaller curtosis)}. In DFT--WBII the density distribution
{has a smaller curtosis} than a Gaussian with $b_4 \approx -0.03$ for the range of densities 1.04 to 1.20
(see panel (c) in Figs.~\ref{fig:rho_r1.04} and \ref{fig:rho_r1.20}). Turning to the asymmetries
 we note that our {\em ansatz}
(Eq.~(\ref{eq:faniso})) describes the data for small distances $r$ very well and starts to deviate
only when the overall density has dropped by a factor $10^4$ compared to the center of the peak. 
This is illustrated in 
panels (b) and (d) in Figs.~\ref{fig:rho_r1.04} and \ref{fig:rho_r1.20} where we compare the fit to
the anisotropic part $f_{\rm aniso}$ to the quotient of
the {density} profile with the purely radial fit, $\rho(\vect r)/(\rho_{\rm G}(r)\; f_{\Delta {\rm  G}}(r)\;)$
(see Eq.~(\ref{eq:faniso_def})). 
The qualitative behavior of the density distribution in the principal lattice directions is
the same for MC and  DFT--WBII, only the magnitude of the leading anisotropy coefficient
$K_4$ is larger in DFT--WBII by about a factor 1.7. The agreement in sign and order of magnitude in $K_4$
with simulations distinguishes fundamental measure theory from the (M)WDA approach where an opposite
sign is obtained \cite{Ohn93}. (Intuitively, the density distribution in [110] direction should be
narrower than in [100] since in [110] direction the next neighbor is closer.)    

\begin{figure}
 \begin{center}
   \epsfig{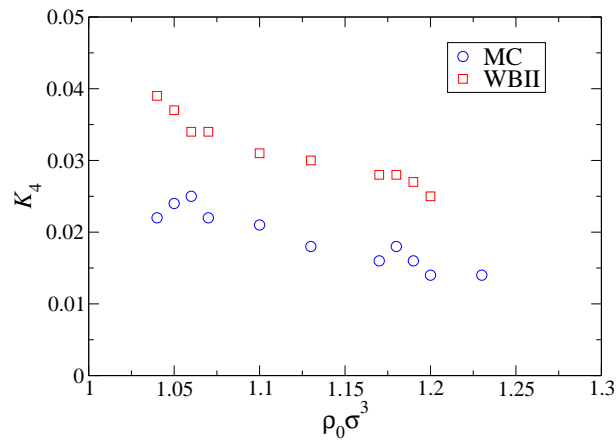} 
 \end{center}
  \caption{\label{fig:k4} Leading anisotropy coefficient $K_4$ vs. bulk density $\rho_0$ as
  obtained from a fit to the density distributions from MC simulations (circles) and 
  from DFT--WBII (squares). 
  }
\end{figure}

In Fig.~\ref{fig:k4} we show the value of $K_4$ for a range of bulk densities from the fits to both 
MC and DFT--WBII results. The scatter in the data is a result of the uncertainty in the fits, but one
can clearly observe a trend to lower $K_4$ for higher density. This would be consistent with
the observation in Ref.~\cite{You74} that towards close--packing the density distribution becomes
Gaussian ($K_4=0$). We note furthermore that we could not extract any meaningful results for the
next--to--leading anisotropy coefficient $K_6$ whose modulus appears to be smaller than $|K_4|$ but the
error estimate is always of about the same magnitude.
{Finally, we remark that our results for $K_4$ are in quantitative agreement to earlier computer simulation 
data published in Ref.~\cite{Ohn93}. }

\section{Discussion and Conclusion}

\label{sec:conclusions}

In this work we have performed a comparative study of the free energies and the 
density distributions in hard sphere crystals using Monte Carlo simulations and 
density functional theory (employing Fundamental Measure functionals). 
Using a recently introduced simulation technique, we could obtain crystal free energies
to a high precision (see Tab.~\ref{tab:Comp}) which are consistent with the most recent parametrizations
of empirical equations of state and allowed us to determine the crystal free energy
in the close--packing limit with a higher accuracy than before (see Eq.~(\ref{eq:fspeedy})).  
The free energies from Fundamental Measure theory are also in good agreement with 
the simulation results and demonstrate the applicability of these functionals 
to the treatment of other problems involving crystallization. The agreement between
FMT and simulations on the level of the free energies is also reflected in the 
density distributions around single lattice sites (see Figs.~\ref{fig:rho_r1.04} and
\ref{fig:rho_r1.20}). Overall, the peak widths and anisotropy signs for different
lattice directions agree, it is found that FMT gives slightly narrower peaks with more
anisotropy than seen in the simulations.   

The deviations we observe between simulation and FMT point to possibilities of
further improvement in the FMT functionals. Tarazona's construction of the
tensor part of these functionals is an approximate representation of the three--cavity overlap
situation (see Fig.~\ref{fig:cav}) which leads to a complicated expression.  
It would be interesting to study the close--packing limit of this expression in a systematic
manner.  

Additionally
we studied theoretically
for the constrained minimization in the unit cell (with particle 
number fixed) the relation of
the respective constrained chemical potential $\mu'$ to Widom's trick in a system with
fixed vacancy concentration $n_{\rm vac}$.
The latter analysis gives a simple relation, $\mu' = {\rm const.} - \ln n_{\rm vac}$
(see Eq.~(\ref{eq:munvacdef})), 
which poses a consistency condition on the corresponding FMT results for $\mu'$. It turns
out that from the three studied variants of FMT, only the White Bear II functional
shows the qualitatively correct behavior whereas the Rosenfeld and the White Bear functional
give qualitatively incorrect results (see Fig.~\ref{fig:vac}). This implies that for further
studies such as the free minimization of the crystal--fluid interface or nucleation processes
only the White Bear II functional is a promising candidate.   

\begin{acknowledgements}
{We thank the DFG (through SFB TR6/N01, SPP 1296/Schi 853/2 {and SPP 1296/Lo 418/13-2}) and the FNR (AFR PHD-09-177) for financial support.} 
\end{acknowledgements}

\end{document}